\documentstyle[tighten,pra,aps,multicol,epsfig]{revtex}

\begin{document}
\draft
\title{Macroscopically distinct quantum superposition states as a bosonic
code for amplitude damping}
\author{ P.T. Cochrane, G.J. Milburn and W.J. Munro}
\address{Department of Physics,
The University of Queensland,
QLD 4072 Australia.}
\date{\today}
\maketitle

\begin{abstract}
We show how macroscopically distinct quantum superposition states
(Schr\"odinger cat states) may be used as logical qubit
encodings for the correction of spontaneous emission errors.
Spontaneous emission causes a bit flip error which is easily
corrected by  a standard error correction circuit. The method works
arbitrarily well
as the distance between the amplitudes of the superposed coherent states
increases.
\end{abstract}

\begin{multicols}{2}
\section{Introduction}
\label{introduction}
In quantum information theory logical states are encoded as two orthogonal
pure states\cite{quantum-info}. The simplest example
is provided by a single two level system. The ground state $|g\rangle$ and
excited state $|e\rangle$ can then encode logical $0$ and
logical $1$ respectively. The ability to form a coherent superposition of
logical states
is why we refer to logical states as qubits rather
than simply as bits\cite{Schumaker}.

Quantum
computation gains its power through the potential ability to
unitarily manipulate a coherent superposition of
large collections of physical systems each encoding a single
qubit\cite{quantum-comp}. There is no fundamental reason
to restrict oneself to physical systems with two dimensional Hilbert spaces
for the encoding. It may be more natural in
some contexts to encode logical states as a superposition over a large
number of basis states. When the system
supporting the qubit encoding is coupled to a perturbing environment an
extra unwanted, and possibly uncontrollable,
unitary interaction is introduced which can appear as an error
in
the encoded information. The coupling
to the logical basis determines the type of logical error. While the
coupling to
the environment is fixed, we are
free to choose how we encode the qubits, hence the choice of basis
for the logical encoding may change the kind of error
introduced. For example, with a single qubit, a bit flip in one
logical encoding basis can appear as a phase flip in another\cite{Braunstein}.
This is relevant, as some kinds of errors are easier to fix than others.
Chuang and Yamamoto\cite{Chuang} recently introduced a qubit
coding for two bosonic modes.

These modes could be two optical modes or two
vibrational modes of a single trapped ion. A
particularly difficult source of error for bosonic modes arises from
exponential decay of the energy. In a single mode, for
example, one could use the ground state and first excited state as the
logical basis. While the ground state is invariant under
decay, the first excited state will `reset' to the ground state in a single
decay event. Such an error can in general be corrected by a
five qubit code\cite{Laflamme98}. However the code of Chuang and
Yamamoto enables a more efficient error
correction.

In this paper we give an example of how a careful choice of the coding
scheme can make a difficult error correction task
simpler. Our example is based on a quantum code for a single bosonic mode
that enables amplitude damping or amplification to be
corrected as a bit flip error.  The code is based on quantum superpositions of
bosonic coherent states, the so called `cat
states'\cite{Knight}. Our coding scheme is not exact for very small amplitude
coherent states, but improves exponentially when
amplitudes are greater than unity. We demonstrate
a completely unitary, adiabatic method to
generate the cat states of our coding scheme.

\section{Cat state encoding for amplitude damping}
Let $|\alpha\rangle$ be a coherent state for a single bosonic degree of
freedom. We then define two orthogonal states as
symmetric and anti-symmetric superposition of coherent states by
\begin{mathletters}
\begin{eqnarray}
|S\rangle & = & {\cal N}_+\left (|\alpha\rangle+|-\alpha\rangle\right)\\
|A\rangle & = & {\cal N}_-\left (|\alpha\rangle-|-\alpha\rangle\right)
\end{eqnarray}
\end{mathletters}
where $\alpha$ is an arbitrary complex number. The normalisation constants
are given by
\begin{equation}
{\cal N}_\pm=\left (2\pm2e^{-2|\alpha|^2}\right )^{-1/2}
\end{equation}
It is easy to verify that the symmetric cat state, $|S\rangle$ contains
only the even energy eigenstates, while the
anti-symmetric cat state $|A\rangle$ contains only the odd energy
eigenstates. This feature is independent of $\alpha$. The
two states are orthogonal and we are led to the following logical encoding
for a single qubit,
\begin{mathletters}
\begin{eqnarray}
|0\rangle_L & = & |S\rangle\\
|1\rangle_L & = & |A\rangle
\end{eqnarray}
\end{mathletters}
Under free dynamics, the coherent state evolves as
$|\alpha(0)e^{-i\omega t}\rangle$, however the two cat states remain
orthogonal and thus the logical encoding of the qubit is
invariant under free dynamics. Therefore we can transform to the interaction
picture rotating at frequency $\omega$.

The amplitude damping model is the standard one for a bosonic mode, of
frequency $\omega$, weakly coupled to a zero temperature
heat bath\cite{Chuang,WallsMil} in the Born and Markov approximation. The
system obeys the following master equation in the
interaction picture
\begin{equation}
\frac{d\rho}{dt}=\frac{\gamma}{2}(2a\rho a^\dagger-a^\dagger a\rho-\rho
a^\dagger a)
\end{equation}
The solution to this equation may be written as\cite{Milburn-ANU}
\begin{equation}
\rho(t)=\sum_{k=0}^\infty\Upsilon_k(t)\rho(0)\Upsilon^\dagger_k(t)
\label{solution}
\end{equation}
with
\begin{equation}
\Upsilon_k(t)=\sum_{n=k}^\infty\sqrt{n\choose
k}(\eta(t))^{(n-k)/2}(1-\eta(t))^{
k/2}|n-k\rangle\langle n|
\end{equation}
and $\eta(t)=e^{-\gamma t}$ is the probability that the state is undecayed
up to time $t$.

Our objective is to correct for at most one decay event over some
characteristic time. In which case we only need to
consider the two terms corresponding to $\Upsilon_0$ and $\Upsilon_1$.
Coherent states remain coherent under amplitude
damping, and in particular we have that,
\begin{mathletters}
\begin{eqnarray}
\Upsilon_0|\alpha\rangle & = &
e^{-(1-\eta)|\alpha|^2/2}|\sqrt{\eta}\alpha\rangle\\
\Upsilon_1|\alpha\rangle & = &
\alpha\sqrt{\eta}e^{-(1-\eta)|\alpha|^2/2}|\sqrt{\eta}\alpha\rangle.
\end{eqnarray}
\end{mathletters}
It is then easy to see that a single decay event will cause an even cat
state to flip to an odd cat state and vice versa. It
is this feature that we are attempting to exploit through our code states,
so that a single decay event will correspond to a
bit flip. A no decay event essentially leaves the state unchanged. These
statements are strictly only true for cats with an
infinitely large coherent amplitude, however we now show that only small
amplitudes are sufficient for practical purposes.

An error correction code must satisfy the following
conditions\cite{Knill}
\begin{mathletters}
\begin{eqnarray}
\mbox{}_L\langle p|\Upsilon_k^\dagger\Upsilon_l|q\rangle_L & = & 0 \ \ \ \
\mbox{for }p\neq q\mbox{ or }k\neq l\\
\mbox{}_L\langle p|\Upsilon_k^\dagger\Upsilon_k|p\rangle_L &=& P_k\ \ \ \ \
\mbox{for } p=0,1
\end{eqnarray}
\end{mathletters}
where $p,q$ are $0$ or $1$, and $P_k$ is a constant that depends only on
$k$. The first equation requires that all erroneous
states are orthogonal and the second requires the probability for each
event (no decay or one decay) to occur to be independent
of the logical state.  It is easy to see that the cat state encoding
satisfies the first condition. The second condition
however requires more careful consideration. Using the conditional states
given above we find that
\begin{mathletters}
\begin{eqnarray}
\frac{\mbox{}_L\langle
0|\Upsilon_0^\dagger\Upsilon_0|0\rangle_L}{\mbox{}_L\langle
1|\Upsilon_0^\dagger\Upsilon_0|1\rangle_L}
& = &
\frac{1+e^{-2\eta\alpha^2}}{1-e^{-2\eta\alpha^2}}\label{eq13}\\
\frac{\mbox{}_L\langle
0|\Upsilon_1^\dagger\Upsilon_1|0\rangle_L}{\mbox{}_L\langle
1|\Upsilon_1^\dagger\Upsilon_1|1\rangle_L}
& = &
\frac{1-e^{-2\eta\alpha^2}}{1+e^{-2\eta\alpha^2}}.
\label{eq14}
\end{eqnarray}
\label{eqs13and14}
\end{mathletters}
Each of these ratios should ideally be unity, but the departure from
ideality is insignificant
even for such a small
value as $\alpha=3$. For example, with $\eta=0.9$ we find
(\ref{eq13}) gives 1.00149 for $\alpha=2$ but for
$\alpha=3$ it gives 1.000000184. While (\ref{eq14}) gives 0.9985079 for
$\alpha=2$ and 0.999999815 for $\alpha=3$. If we
increase the amplitude to $\alpha=5$ the departure from ideality is
undetectable. Therefore the logical qubits are
encoded in a manner that enables amplitude decay to be corrected to any
desired degree of precision.

We can see that after many spontaneous emission events the amplitude
will eventually decay away to zero.  If the coherent amplitude is too 
small then the ratios (\ref{eq13}) and (\ref{eq14}) will deviate 
significantly from unity.  It is therefore
necessary to have a sufficient initial amplitude to allow computation
for a reasonable amount of time and to know when it is prudent to 
reset the states.

It is possible to determine the time scale over which the states will be 
useful
by considering
the ratio (\ref{eq13}).  This ratio should not be significantly 
different from 1 for the encoding to work, so we allow the difference to 
be no greater than a small
tolerance.  The term 
responsible for any deviation of the ratio is $\exp(-2\eta\alpha^{2})$ 
which we desire
to be small
enough such that the ratio is within tolerance.
This implies that $\eta\alpha^{2}$ has to be greater than some limiting 
value
determined by the tolerance, below which the state must be reset.  
Therefore, given a certain error rate, initial coherent amplitude and 
desired tolerance we have sufficient information to calculate the 
time available for computation before reset.

\section{Logical operations on cat states}
A logical encoding is useless if we cannot implement one and two qubit
operations on the encoded states. We now show how this
can be done for the cat state encoding defined above. The particular form
of qubit operations depends upon the particular
physical realisation of the bosonic mode. For the purposes of illustration
we simply postulate particular bosonic interactions
to achieve the required gate operations.  We will show that the
Hadamard transform may be implemented by
simple displacement of a single bosonic mode,
while the two qubit operation may be realised by a mutual phase shift
interaction term which commutes with the number operator
of each bosonic mode.

If the bosonic mode is subject  to a classical driving force the
Hamiltonian describing this process in the interaction picture is
\begin{equation}
H_D=\hbar(\beta a^\dagger+\beta^* a)
\end{equation}
where $\beta$ is the complex amplitude of the driving force. Let us now
choose $\beta$ as real (in general we choose $\beta$ to be $\pi/2$ out
of phase with $\alpha$).

For a given cat state amplitude we can choose the
driving amplitude such that
\begin{equation}
\theta=\alpha\beta t
\end{equation}
where $t$ is the length of time the driving force is applied.

If the even cat state (encoding $|0\rangle_L$) is driven we find that
\begin{equation}
e^{-iH_D t/\hbar}|0\rangle_L=\cos\theta
|0\rangle_L-i\sin\theta|1\rangle_L
\end{equation}

A displacement of this kind shifts the ``cat'' very slightly by an amount
$\beta$ in a direction orthogonal to the orientation of the cat state in
phase space.  The transformation is approximately equivalent to a Hadamard
transform of
the single logical qubit when $\theta=\pi/4$ (in the limit of large $\alpha$
and small $\beta$) and
will suffice as a universal one-qubit gate. We will refer to this as a H-gate.

The simplest way to realise a two qubit universal gate is via the two mode
interaction Hamiltonian
\begin{equation}
H_P=\hbar\chi a^\dagger a b^\dagger b
\end{equation}
where $a,b$ represent the mode amplitude operators for the two bosonic
modes of interest. We choose the interaction time
$t$ such that $\chi t = \pi$. As the $|0\rangle_L$ only has even bosonic
number while $|1\rangle_L$ only has odd bosonic
number, we find that the interaction leaves the states,
$|0\rangle_{La}|0\rangle_{Lb},\
|0\rangle_{La}|1\rangle_{Lb},\ |1\rangle_{La}|0\rangle_{Lb}$
unchanged, but the state
in which both modes encode a $|1\rangle_L$ transforms
as
\begin{equation}
e^{-i\pi a^\dagger ab^\dagger b
}|1\rangle_{La}|1\rangle_{Lb}=-|1\rangle_{La}|1\rangle_{Lb}
\end{equation}
This kind of conditional phase shift operation suffices for a universal two
qubit gate. We will refer to this as a P-gate.

Using the one and two qubit gates described above, we can
construct a controlled-not (CN) gate. Let mode $a$
code the control bit and mode
$b$ code the target bit.  A CN gate is then made by applying a
H-gate to the target, then coupling the target and the
control by a P-gate and finally applying another H-gate to the target.

We have shown that simple one mode and two mode transformations may be
used to construct universal computational gates
for a cat state logical encoding  of bosonic systems. Amplitude damping
appears as a simple bit-flip error in this encoding and
thus a three qubit code can be used to correct it. This
leads to relatively simple fault tolerant
implementations of the gate operations described above using three coupled
bosonic modes.

\section{Unitary construction of cat state encoding}
The cat state encoding described in this paper will be of little use if we
cannot encode our logical bits by unitary
transformations. Unfortunately all previous schemes to generate cat states
are based on an entanglement between a bosonic mode
and a two level atom and require a measurement readout\cite{WineReview}.
The cat state
produced is conditional on the two, mutually exclusive,
results of this measurement, and  we are equally likely to get an even cat
as an odd cat. This method of encoding would randomly
assign logical bits and is of little practical use. We now describe a
unitary, although adiabatic, method to generate the two
kinds of cat state used to encode the qubits.

Consider the Hamiltonian,
\begin{equation}
H_{NL}=\hbar\chi (a^\dagger)^2 a^2
\label{kerr}
\end{equation}
which could describe a Kerr nonlinearity for an optical bosonic
mode or the  self-interaction of a single
trapped ion driven at the carrier frequency\cite{Filho96}, in which case
$\chi$ is proportional to the fourth power of the
Lamb-Dicke parameter. The Hamiltonian in (\ref{kerr}) has two degenerate
ground states which are the ground state,
$|0\rangle$ and first excited state, $|1\rangle$ of a single bosonic mode. In
both cases the eigenvalue is zero, however each
of these ground states is distinguished by the parity operator, where
$|0\rangle$ is even and $|1\rangle$ is odd. If we now
consider the Hamiltonian
\begin{equation}
H_C=H_{NL}-\hbar\kappa(a^2+(a^\dagger)^2)
\end{equation}
with $\kappa\ \geq\ 0$.
Noting that the cat states $|0\rangle_L,\ |1\rangle_L$ are eigenstates of
$a^2$, it is easy to see that these same states are
degenerate eigenstates of $H_C$ when
$\alpha = \sqrt{\kappa/\chi}$ 
and the eigenvalue is $-\hbar\kappa^2/\chi$. While the cat states are
degenerate eigenstates of $H_C$ they are distinguished by
their parity. The adiabatic theorem now enables us to predict that the even
or odd initial eigenstates of $H_{NL}$,
$|0\rangle,\ |1\rangle$, will evolve respectively into the even or odd
eigenstates, $|0\rangle_L,\ |1\rangle_L$ of $H_C$ as we
slowly turn on $\kappa$ from zero to a final target value. Thus we have a
unitary method to code either a logical zero or
logical one as a cat state by choosing to start from a bosonic ground state
or a bosonic first excited state.

The adiabatic theorem is exact only in the case of infinite slowness, which
is of little use for logical encoding in quantum
computation, so what matters is how well we can do in practise. To test this
we consider two different ways to vary $\kappa$ in
time; linear and nonlinear.

The linear variation considered here consists of simply increasing
$\kappa$ according to $\kappa=t$.  The function
$\kappa = k_0 \tanh^2 (\lambda t)$ was
used in the nonlinear case due to the advantageous shape of the
$\tanh^2$ function.

Figure \ref{figtwo} illustrates the fidelity versus time variation of
the state
$|\psi\rangle$ starting from the $|1\rangle$ Fock state with respect to the
equivalent cat
state
of mean photon number $\alpha = \sqrt{\kappa/\chi}$.  The fidelity is
measured as the modulus squared of the dot product of the evolving
state with the cat state.
The notable features of Figure \ref{figtwo} are the
fidelity oscillations, the ``steady-state'' fidelity
and the relative characteristics of the linear
and nonlinear methods of varying $\kappa$.

The fidelity oscillations result from carrying out the adiabatic
evolution faster than as required for exactness by the adiabatic
theorem.  As the system evolves from $t = 0$ the fidelity will tend away
from unity as
the state $|\psi\rangle$ evolves away from the relevant cat state.
Continued evolution eventually causes $|\psi\rangle$ to more closely
resemble the
equivalent cat state with the fidelity increasing accordingly.  The
retreat and approach of the evolving state with respect to the
cat state causes the oscillations seen in Figure \ref{figtwo}.

The oscillations are damped by $\kappa$ until a
``steady state'' is reached with constant fidelity.
The ``steady state'' fidelity is determined by how quickly $\kappa$ is
increased from $t=0$; a slower initial increase implies a greater
final fidelity.  Hence there are two effects occurring with $\kappa$:
as $\kappa$ increases, oscillations in fidelity are suppressed, and
the faster $\kappa$ is increased initially, the lower the steady state
fidelity of $|\psi\rangle$.

For this adiabatic process to be useful we have two aims; a steady state in
a reasonable amount of time, and a state $|\psi\rangle$ as close as
possible to
the desired final state.  For linear $\kappa$ these are
complementary,
although for nonlinear $\kappa$ we can choose a function which can
achieve both aims, hence the use of the $\tanh^2$ function (s-curve).
The s-curve has the properties that it starts
slowly, thus giving a high final fidelity, and later damps the system
very quickly to give a useful final state in a reasonable amount of
time.  If the variables $k_0$ and $\lambda$ are chosen carefully then
it is possible to obtain a fidelity of almost unity in a
usefully short time.

We thus conclude that the unitary logical encoding in terms of cat states may
be performed with almost arbitrary accuracy using this adiabatic
method.

\section{Discussion and Conclusion}
We have shown that the even and odd cats states may be used as a robust
qubit encoding for a single bosonic mode subject to
amplitude damping. A single decay event will then appear as a simple bit
flip error. We have also shown how the states may be
prepared unitarily and how one qubit and two qubit universal quantum gates
may be realised. We now turn to an assessment of how
practical the scheme is for present technology. To be specific we will
consider the case in which the bosonic mode is the
centre-of-mass vibrational state of a single trapped ion. Cat states have
been produced in these systems using a conditional
measurement scheme\cite{wine}.

Given a cat state it is straightforward to protect it against decay using
two additional qubits. These could be the
electronic states of two ions in the trap. The error correction circuit for
a bit flip is well known and is given in
Figure \ref{fig1}. To implement the gate we need to implement a CN gate
between the vibrational state and the electronic
states of the two ions. Following de Matos Filho and Vogel\cite{Filho96} we
consider an ion trapped
at an antinode  of an optical standing wave tuned to the atomic frequency;
the carrier
frequency.  In an interaction picture at frequency $\nu$ the interaction
Hamiltonian is,
\begin{equation}
H_I=-\hbar\Omega\eta^2 a^\dagger a\sigma_x +
\hbar\frac{\Omega\eta^4}{4}(a^\dagger)^2a^2\sigma_x
\label{ham}
\end{equation}
where $\Omega$ is the Rabi frequency and $\eta$ is the Lamb-Dicke
parameter. The first term in this expression suffices to build a CN gate
between the cat state and the electronic state. If
we choose the interaction time appropriately we can apply the transformation
\begin{equation}
U=\exp(-i\pi a^\dagger a \sigma_x)
\label{ioncn}
\end{equation}
When this acts on an even cat state it corresponds to the identity on the
electronic system. When it acts on an odd cat state
it corresponds to a $\pi$ pulse in the electronic system. If we
code our electronic qubits as $|g\rangle_1\rightarrow
|0\rangle_i$ and $|e\rangle_1\rightarrow
|1\rangle_i$. The unitary interaction in (\ref{ioncn}) will effect a CN
gate with the bosonic mode acting as the control
and the electronic mode acting as the target. Thus joint excitation on the
carrier frequency of the two ion system will
produce the double CN gate in the first part of Figure \ref{fig1}. The
final double CN gate in which the vibrational mode
becomes the target can easily be produced with the same Hamiltonian with
H-gates either side.
This procedure would enable a cat state, once produced, to be protected
from single decay events.

\begin{acknowledgements}
GJM would like to thank the Benasque Centre for Physics for support of the
visit at which this paper was written.  PTC acknowledges the financial
support of the Centre for Laser Science and the University of Queensland
Postgraduate Research
Scholarship.
\end{acknowledgements}

\begin{center}
\begin{figure}[!ht]
\narrowtext
\epsfig{figure=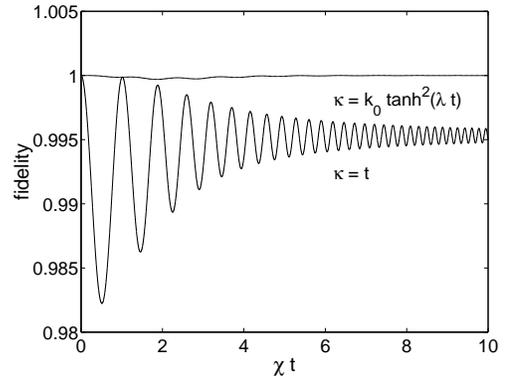,width=65mm}
\caption{Fidelity-time evolution, $\psi$ starting from $|1\rangle$.}
\label{figtwo}
\end{figure}

\begin{figure}[!ht]
\narrowtext
\epsfig{figure=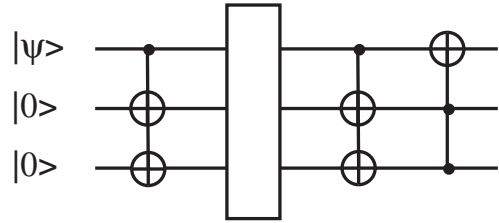,width=65mm}
\caption{3-qubit circuit to correct bit-flip errors.}
\label{fig1}
\end{figure}
\end{center}

\end{multicols}
\end{document}